\documentstyle[preprint,aps,psfig]{revtex}
\draft
\tightenlines
\title{Persistence and Life Time Distribution in Coarsening Phenomena}
\author{V. Sridhar,  K. P. N. Murthy and M. C. Valsakumar}
\address{Theoretical Studies Section, 
         Materials Science Division\\
         Indira Gandhi Centre for Atomic Research\\
         Kalpakkam 603 102\\
         Tamilnadu, India}
\begin{document}
\maketitle
\begin{abstract}
We investigate the life time distribution  $P(\tau , t)$ in one dimensional
and two dimensional coarsening processes modelled by Ising-Glauber dynamics at
zero temperature. The life time $\tau$  is defined as the time that
elapses between two successive
flips in the time interval $(0,t)$ or between the last flip and the observation time 
$t$.  We calculate $P(\tau , t)$ averaged over all the spins
in the system and  over 
several initial disorder configurations. We find that  asymptotically the 
life time distribution obeys a scaling ansatz: $P(\tau\ ,\ t)= t^{-1}\phi (\xi)$,
where $\xi=\tau /t$. The scaling function $\phi (\xi)$ is singular at $\xi =0$ and $1$, 
mainly due to slow dynamics and persistence. An independent life time model 
where the life times are sampled from a distribution with power law tail is presented, 
which predicts analytically the qualitative features of the scaling funtion.  The need for going 
beyond the independent life time models for predicting the scaling function for the 
Ising-Glauber systems is indicated. 
\end{abstract}
\bigskip
Coarsening phenomenon\cite{ajb},
is a
simple example of a dynamical process which is slow, and which becomes 
slower  as time proceeds. 
This phenomenon is found in several nonequilibrium systems, {\it e.g.}
phase separation in binary alloys, grain growth, 
growth of soap bubbles, magnetic bubbles {\it etc.} 
A striking feature of the coarsening phenomenon is the dynamical scale 
invariance. The domain structure at different times are statistically
similar to each other but for a rescaling length  $L(t)$. 
The rescaling length  can be taken as 
the typical 
linear size of a domain and it increases with time as $t^{1/z}$.
It was soon realized that the dynamical scaling exponent $z$ is not 
adequate to describe  completely  the coarsening dynamics, see below. 

An interesting question asked in this context concerns the persistence 
probability $P_0 (t)$, that a local order parameter 
does not change its sign  until the observation 
time $t$. In other words,
$P_0 (t)$ refers to the fraction of the total volume that remains
unswept by the domain walls until time $t$,  in the context of
three dimensional coarsening phenomenon.
This quantity, exhibits, asymptotically, a power law 
decay,  $P_0 (t) \sim t^{-\theta}$, where  
$\theta$ is called the persistence exponent, and it is 
 independent of $z$, the dynamical scaling exponent. 

The persistence phenomenon has a long history. The earliest question on 
persistence was perhaps asked and answered by 
W. A. Whitworth in 1878 and later by 
J. Bertrand in 1887, see Feller\cite{Feller}. The question relates 
to two canditates $P$ and $Q$, who poll $p$ votes 
and $q$ votes respectively, with 
canditate $P$ winning the election by a margin of $x=p-q$ votes. Let $n=p+q$
denote the total number of votes polled and there are no invalid votes! 
The ballot problem 
consists of finding the probability that throughout the counting process
the canditate $P$ leads.  
In the language of random walks, this persistence probability equals the ratio
of the number of walks that start at origin, remain above origin and 
eventually reach the site $x$ after $n$ steps to the total number of walks 
that start at origin and reach site $x$ after $n$ steps. 
   A simple application of reflection principle yields the persistence 
probability as $x/n$, see Feller\cite{Feller}, which equals
the fractional excess votes polled by $P$ over $Q$.

The next question on persistence arose in the context of simple 
random walks on a one dimensional lattice. 
Let us consider the walks that 
start at origin and take $2n$ steps. The total number of random walks is thus
$2^{2n}$ and we consider that all these walks to be equally probable. Let us 
collect the fraction of random walks that visit origin for the last 
time at step $2k$. This fraction is given by,
\begin{equation}
P( 2k,2n)=
{{ (2k)!}\over{ k! k!}}\  {{(2n-2k)!}\over{(n-k)!(n-k)!}}\ {{1}\over{2^{2n}}}
\end{equation}
$P(2k,2n)$, given above, is called the discrete 
arc-sine distribution of order $n$. It is $\cup $-shaped; it is maximum at 
either ends of the support {\it i.e.} at $k=0$ and at $k=n$. It is symmetric.
Let us define a  scaling variable $\xi _k = k/n$ and express 
$P(2k,2n) = n^{-1}\phi (\xi _k )  $. 
The continuous arc-sine distribution,
\begin{equation}\label{arcsine}
f (\xi) = {{1}\over{\pi}}\ {{1}\over{\xi ^{1/2} (1-\xi )^{1/2 }}}
\end{equation}
provides a very good fit to the discrete scaling function $\phi (\xi _k )$,
 especially 
for large $n$ and $k$ not too close to $0$ or $n$.  
Eq. (\ref{arcsine}) 
was first derived by P. L\'evy\cite{Levy} for the Brownian motion,
which is a continuum  version of the random walk problem considered above. 
The persistence probability distribution as given by the 
arc-sine law is a surprising result, since contrary to 
intuition, it says that in a duration of time $t$, 
the  probability for the last zero crossing  
is maximum at time zero and at time $t$.  
In other words, the probability is highest for the Brownian particle to be left
or right of the origin persistently. 
The $\cup $ shaped 
arc-sine distribution with singularities at either ends of the 
support is the fore-runner to  all the surprising results 
found in the recent times in the context of persistence in 
Ising and Potts spin systems, coarsening systems {\it etc.}, including the 
ones presented in this paper.
For a review of the persistence phenomenon see Majumdar\cite{snm}.

The phenomenon
of persistence has been studied extensively both theoretically
\cite{bd,theory} and experimentally\cite{experiment} in the recent times.
 Also the notion of
persistence of local order 
has been extended to persistence of global order, of domains,
and of patterns\cite{extensions}.
We propose here yet another generalization of the phenomenon of 
persistence for characterizing the slow dynamics of the 
coarsening process. 

Accordingly, we investigate  first the one dimensional zero temperature Ising-Glauber dynamics 
for the coarsening phenomenon. Consider a one dimensional array of  $N$ Ising spins
$\left\{ S_i \ =\ \pm 1\ :\ i=1,N\right\}$.  
The spin system is prepared in a homogeneous, disordered, high temperature
phase. Operationally this means that we assign to  each spin a value of $+1$ or $-1$ 
independently and with 
equal probability. A spin interacts with its nearest neighbours only and the 
interaction is ferromagnetic. The Hamiltonian is given by,
$
H = -J \sum_{\left\langle i,j\right\rangle} S_i S_j
$, 
where the sum runs over nearest neighbour pairs and $J$ measures the strength of ferromagnetic
interaction. We set $J=1$, without  loss of generality. 
Periodic boundary condition is imposed. At time $t=0$,
we quench the system to zero temperature. Upon the temperature quench, the system does not order 
instantaneously or even immediately. Instead, domains of equilibrium broken-symmetry 
phases form, grow and the domain structure coarsens.

The dynamics is simulated as follows. A spin is selected randomly. It is always flipped
if the resulting configuration has lower energy; the spin is 
never  flipped if the 
energy is raised; the spin is flipped with probability half if the change in energy is zero. 
A set 
of $N$ consecutive spin-flip attempts constitutes a Monte Carlo
time Step (MCS) which sets the unit of time. Spin-flips occur relatively
often in the initial times since there would exist 
a large number of lattice sites where the spin flips are energetically
favourable. As time proceeds, domains of up spins and down spins form,
coalesce and grow.  Since only the spins in the domain boundaries 
can flip, the dynamics slows down. The  dynamics becomes slower and slower
since the number of domains decreases with time.

We say that when a spin flips, it dies and  is reborn instantaneously  
in the flipped orientation. We denote by $\tau$ the time that elapses
between two consecutive spin flips, and call it the life time. Thus 
a single spin can have many lives of possibly different life times. 
$\tau$ is a random variable and we are interested in its 
distribution obtained from an ensemble of all the lives of all the spins
in the system. Further we average the life time distribution  
over  several initial
disorder configurations. Suppose during the time interval
$(0,t)$, a spin flips for the last time at say $t_L$ and does not
flip  until time $t$. Then we take $t-t_L$ as the last life time
of the spin. The average life time distribution is obviously dependent
on the observation time $t$. We denote this by $P(\tau , t)$.
We define a scaling variable $\xi = 
\tau / t$, and make a scaling ansatz that for $t\to\infty$,
$P(\tau , t)
\sim t^{-1}\phi (\xi )$.
Fig. 1 depicts $tP(\tau , t)$ {\it vs} $\xi$ for several values of $t$ on a semi-log graph.
 We find that the 
life time distributions for large  $t$, collapse reasonably well  
establishing the validity of the scaling ansatz. 
In Fig. 1, we see that the collpase occurs for $t\ge 4000$ MCS. We have also shown
the life time distributions for $t=500$ and $1000$, and they deviate, though not 
considerably, 
from the scaling curve. 
One can see from Fig. 1 that the scaling curve is $\cup$ shaped  and exhibits 
singularities at either ends of its support $(0,1)$, a feature 
observed in the arc-sine law arising in the context of Brownian
motion.  

Consider now the distribution $P(\tau  , t)$, in the limit $\tau\to t$. 
This limit requires that the spin never flips during the 
time interval $(0,t)$. Let us define $Q(t) = P(\tau =t , t)$.
Clearly $Q(t)$ is the same as the persistence probability $P_0 (t)$
but for a time dependent normalization $N(t)$, defined as the average
number of lives per spin. In other words, $Q(t) = P_0 (t)/N(t)$.
Fig. 2  depicts $Q(t)$ {\it vs.} $t$ on a log-log graph. The 
points fall on a straight line, and  a linear least square estimate gives 
the
slope as  $-\theta _L = -0.87$, implying that $Q(t)
\sim t^{-\theta _L}$. Fig. 3  depicts a similar plot for the 
average number of lives per spin. We find that $N(t)\sim t^{\theta _{N}}$,
with $\theta _{N}$ estimated as $0.49$. The exponent $\theta _N$ 
lends itself to a simple explanation. It is related to the 
number of zero crossing of a Brownian particle in a duration of time $t$,
which goes \cite{Feller} as $t^{1/2}$.  From these two exponents 
$\theta _L$ and $\theta_N$, we can  now calculate the 
standard persistence exponent as $\theta = \theta _L - \theta _{N}= 0.38$, a
result consistent  with earlier results 
\cite{bd}. 
The singularities at $\xi = 0$ and at $\xi =1$, of the scaling 
curve $\phi (\xi)$, can be understood as follows. 
In the discussions below, we always take the limit $\tau\to\infty$, 
$t\to\infty$ such that $\xi=\tau/t$ is finite. Let us first consider the singularity at
$\xi =0$. We notice that only those spins in the domain boundaries can flip. 
As time progresses, the number of domains decreases as $t^{-1/2}$. Also once a spin 
flips the probability it flips next in time $\tau$ is related to the 
first return to origin of a Brownian particle, and is proportional to $\tau^{-1/2}$.
Thus we get,
$P (\tau , t)= t^{-1}\phi (\xi) \sim t^{-1/2}\tau ^{-1/2}$.
Hence in the limit $\xi\to 0$, we have,
 $\phi (\xi) \sim  \xi ^{-1/2}$.
For understanding the other singularity (at $\xi =1$), we use the 
notion of persistence. The limit $\xi\to 1$ implies $\tau\to t $. 
In this limit,
$P(\tau , t)=
t^{-1}\phi (
\xi) \sim t^{-\theta _L}$.
Hence $\phi (\xi ) \sim t^{-\theta_L +1}$;
also $t$
is of the order of $(1-\xi)^{-1}$. Thus we find that
in the limit $\xi\to 1$,
$\phi (\xi) \sim  (1-\xi )^{-(1-\theta_L)}$.

Next we consider the two dimensional zero temperature 
Ising-Glauber model for coarsening phenomenon. 
In Fig. 4 we depict the distribution of the life time in the 
scaling form. The scaling curve looks qualitatively the same as the 
one found for the one dimensional case. The data collapse 
for large observation times $t$ is very clear. In Fig. 5  we depict 
$Q (t)$ {\it vs.} $t$ on a log-log graph. The data points fall on a 
straight line showing that $Q (t) \sim t^{-\theta _L }$ and the 
exponent $\theta _L = 0.58$. Fig. 6  depicts the average number of 
lives per spin, $N(t)$ {\it vs. } $t$ on a log-log graph, and  we find that 
$N(t) \sim t^{\theta _{N}}$. The exponent $\theta _{N} = 0.36$.
Unlike the one dimensional Ising-Glauber dynamics, this exponent 
does not have a simple analytical explanation. It has to be calculated 
only numerically, atleast as of now. 
From the numerical estimates of $\theta_L$ and $\theta _N$, 
we can calculate the standard persistence exponent as 
$\theta = \theta _L - \theta _{N} = 0.22$, a result which matches with an 
earlier finding\cite{bd}.

At this stage, one can think of 
a simple model for the coarsening dynamics, in terms of sampling  the life times independently 
and randomly from a distribution, which we denote by $\rho _1 (\tau)$. The persistence 
exponent appears in this model as a parameter and our interest is to investigate the 
nature of the $\cup$ shaped life time distribution. 
 In such an approach,
we neglect the spin-spin correlations that are present in the Ising model.  
The hope is that such simple 
models would help us understand the nature of the persistent events 
underlying these distributions and  help us in investigtating  
persistence in extended nonequilibrium statistical mechanical models. 
The distribution $\rho _1 (\tau)$ is taken as power-law tailed.
The  reason is clear, if we consider that within the scope of independent 
life time model, the persistence probability is 
\begin{equation}
P_0 (t)\   {}^{\ \ \sim}_{t\to\infty}\ 
t^{-\theta_L } = \int _t ^{\infty} \rho _1 (\tau ) d\tau ,
\end{equation}
which implies that for large $\tau$, we have,
\begin{eqnarray}\label{rho1}
\rho_1 (\tau ) & \sim  & {{\theta_L }\over{\tau ^{1+\theta_L }}} .
\end{eqnarray}
We take the above as the distribution of a single life time. Also
 $1\ \le \ \tau\ \le\  \infty$ and $\theta _L \ < \ 1.$
We set the lower limit of $\tau$ at unity for ensuring normalization.
Let $\rho_m (\tau) $ denote the distribution of the sum of $m$ independent
realizations of $\tau$. It is easily seen that,
\begin{equation}
\rho _m (\tau ) =
   \theta_{L} ^{-1/m} \rho_1 \left( {{\tau}\over{\theta_{L}^{1/m}}}\right).
\end{equation}
With this model for  the basic life time distribution,  we are now ready to
derive the scaling function. 
We define $t_n = \sum_i ^n \tau _i$, the sum of $n$ independent 
realizations of the random variable $\tau$, sampled from the distribution
$\rho_1 (\tau)$. We define $\xi = \tau / t_n$.  Formally we have,
\begin{equation}
\phi (\xi) = \int d\tau\int dt_n \ \delta (\xi -\tau/t_n )\ \rho_1 (\tau)\ \rho_{n-1}(t_n - \tau) .
\end{equation}
Noting that
$
\delta (\xi -\tau /t_n )=t_n \delta (\tau -\xi t_n) ,
$
 and carrying out the inegration over $\tau$, we get,
\begin{equation}\label{rho}
\phi(\xi)=\int_{\beta}^{\infty} dt_n t_n \rho_1 (\xi t_n) (n-1)^{-1/\theta_L }\rho_{1}
 \left( {{ t_{n}\left( 1-\xi\right) }\over{ (n-1)^{1/\theta_L }}}\right)
\end{equation}
where $\beta$ is the lower limit of the integration, obtained
suitably, see below. In the above we substitute for $\rho _1 (\cdot )$
given by Eq. (\ref{rho1}) and carry out the integration. We get,
\begin{equation}
\phi (\xi ) = {{ (n-1)\theta_L }\over{ 2\xi ^{1+\theta_L } (1-\xi)^{1+\theta_L }}} \beta
^{-
2\theta_L }
\end{equation}
We note that the argument of the function $\rho_1 (\cdot)$ occuring in Eq. (\ref{rho})
must be greater than unity. This requirement gives rise to the following inequalities,
\begin{eqnarray}
\xi t_n &>& 1\\
{{t_n (1-\xi)}\over{ (n-1)^{1/\theta_L } }}  & > & 1
\end{eqnarray}
From the above we get $\beta$ as,
\begin{eqnarray}
\beta = max \left\{ {{1}\over{\xi}} ,
{{ (n-1)^{1/\theta_L }}\over{1-\xi }} \right\}
\end{eqnarray}
Finally we get,
\begin{eqnarray}
\phi (\xi ) = \cases{ {{(n-1)\theta_L }\over{ 2 (1-\xi)^{1+\theta_L }}} 
                 \ {{1}\over{  \xi ^{1-\theta_L }}}
                    &\   for   $0 < \xi < \xi ^{\star}  $\cr
          \         & \                                  \cr
                   {{\theta_L }\over{2(n-1)\xi ^{1+\theta _L   }}}\   {{1}\over {  
                   (1-\xi)^{1-\theta_L } }}
                    &\     for     $\xi ^{\star} < \xi  <  1 $\cr  }
\end{eqnarray}
where,
\begin{equation}
\xi ^{\star} = {{1}\over{1+(n-1)^{1/\theta_L } }}
\end{equation}
First observation we
make is that the scaling function is singular at the either ends of the
support, a feature observed  in the one and two dimensional 
Ising-Glauber models presented in this paper.  
We find that the  scaling function
is explicitly a function of $n$,  the number of lives of a spin.
The exponents characterizing the singularities at $\xi =0$ and 
at $\xi =1$ are both equal. However, for the Ising-Glauber 
models, the exponents are different. Hence we may need to go beyond
the independent life time models, for making contact with the 
observations on Ising-Glauber dynamics. A correlated life time 
model could prove useful in this context. Work in this direction is in progress and 
will be reported soon\cite{smv}

We dedicate this work to the memory of Klaus W. Kehr. 

\newpage
\begin{figure}
\centerline{
\psfig{figure=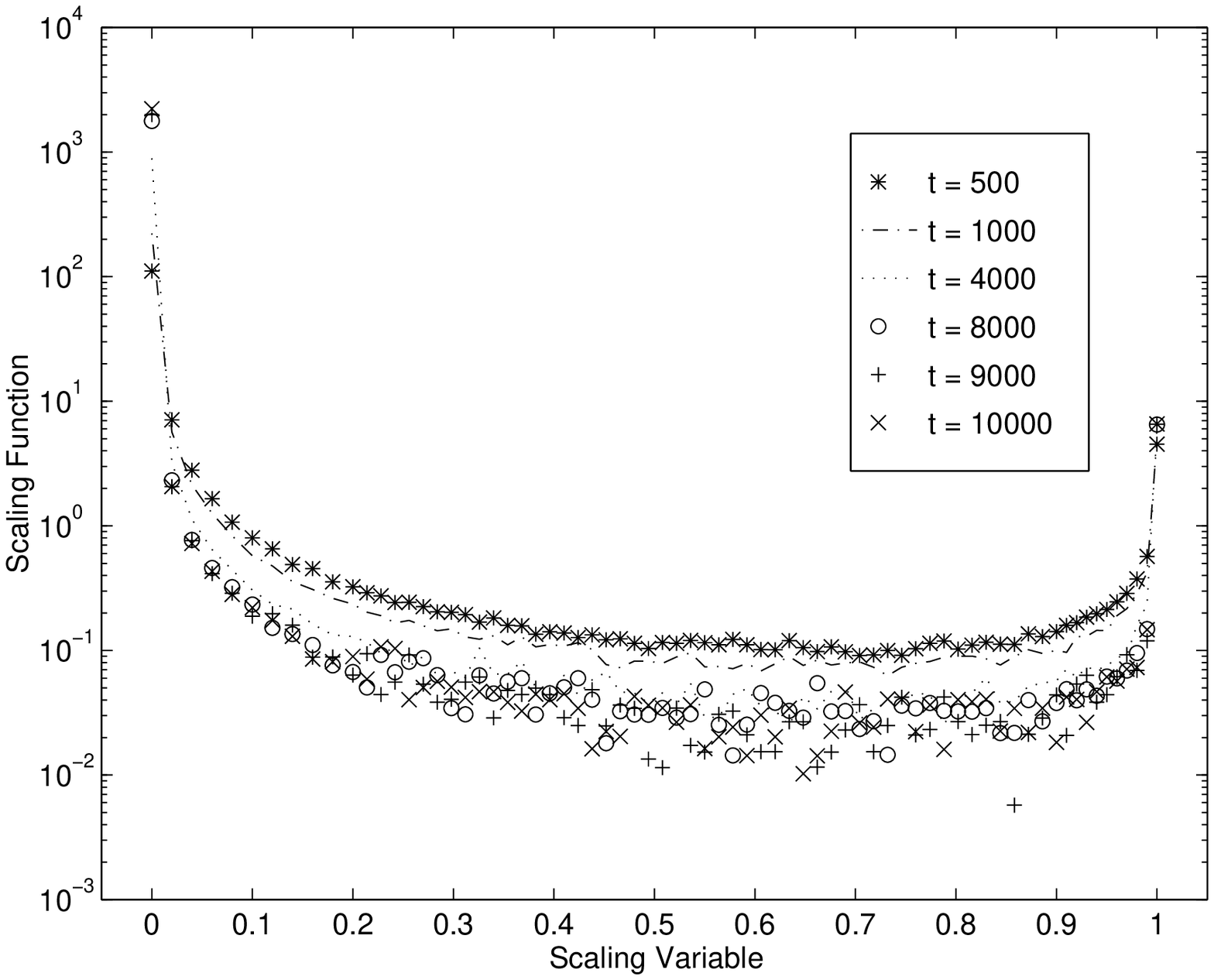}}
\caption{\protect The life time distribution function plotted against the scaling variable,
        for the zero temperature one dimensional Ising-Glauber coarsening.}
\end{figure}
\newpage
\begin{figure}
\centerline{
\psfig{figure=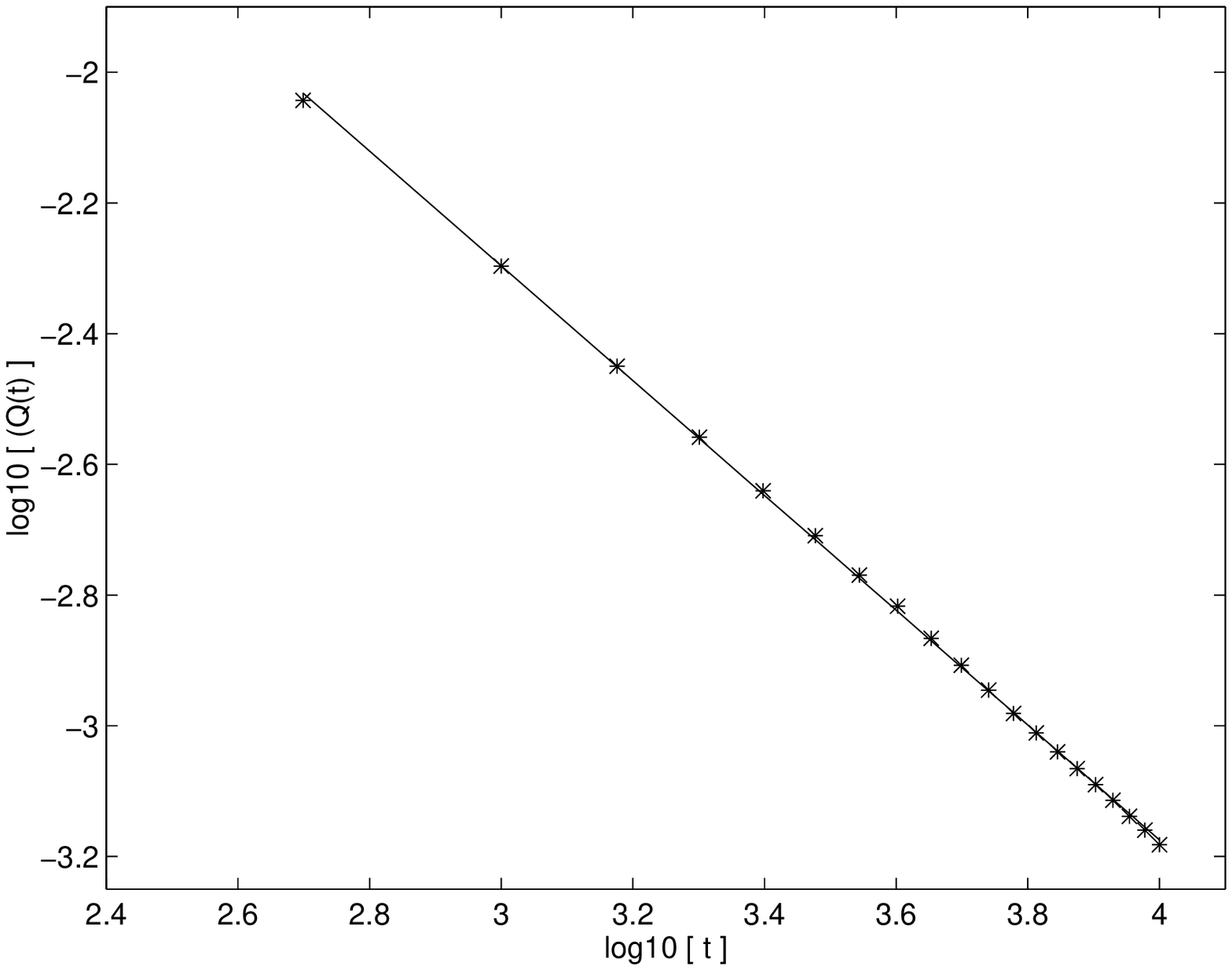}}
\caption{\protect  The persistance probability $Q(t)$ {\it vs.} $t$, for the zero temperature
           one dimensional Ising-Glauber coarsening. The solid line
           corresponds to the linear least square fit. The slope is
           $-0.8744$
}
\end{figure}
\newpage
\begin{figure}
\centerline{
\psfig{figure=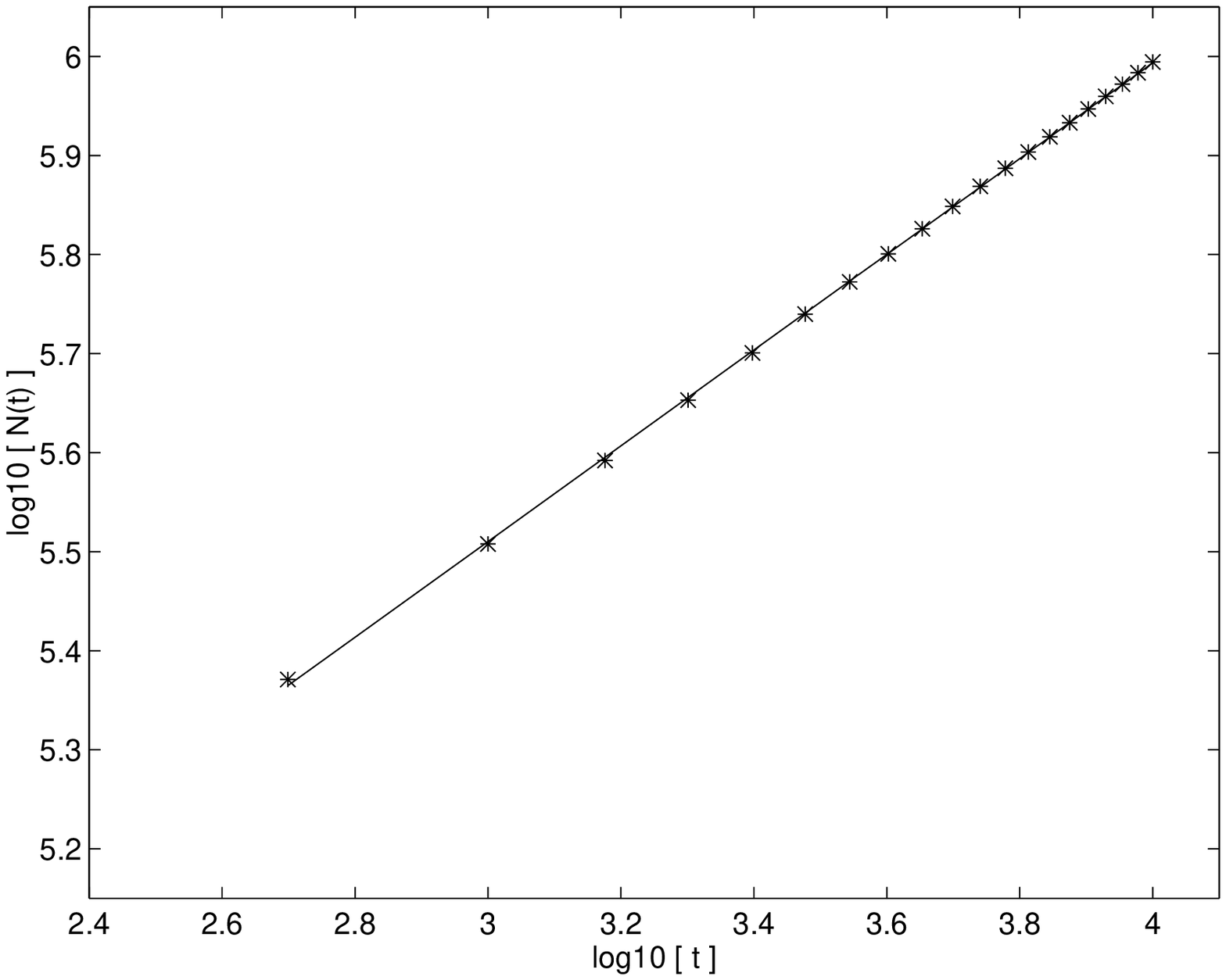}}
\caption{\protect he average number of lives per spin, for the zero temperature
           one dimensional Ising-Glauber coarsening.
            The solid line is the linear least square fit.
            The slope is $0.4853$
}
\end{figure}
\newpage
\begin{figure}
\centerline{
\psfig{figure=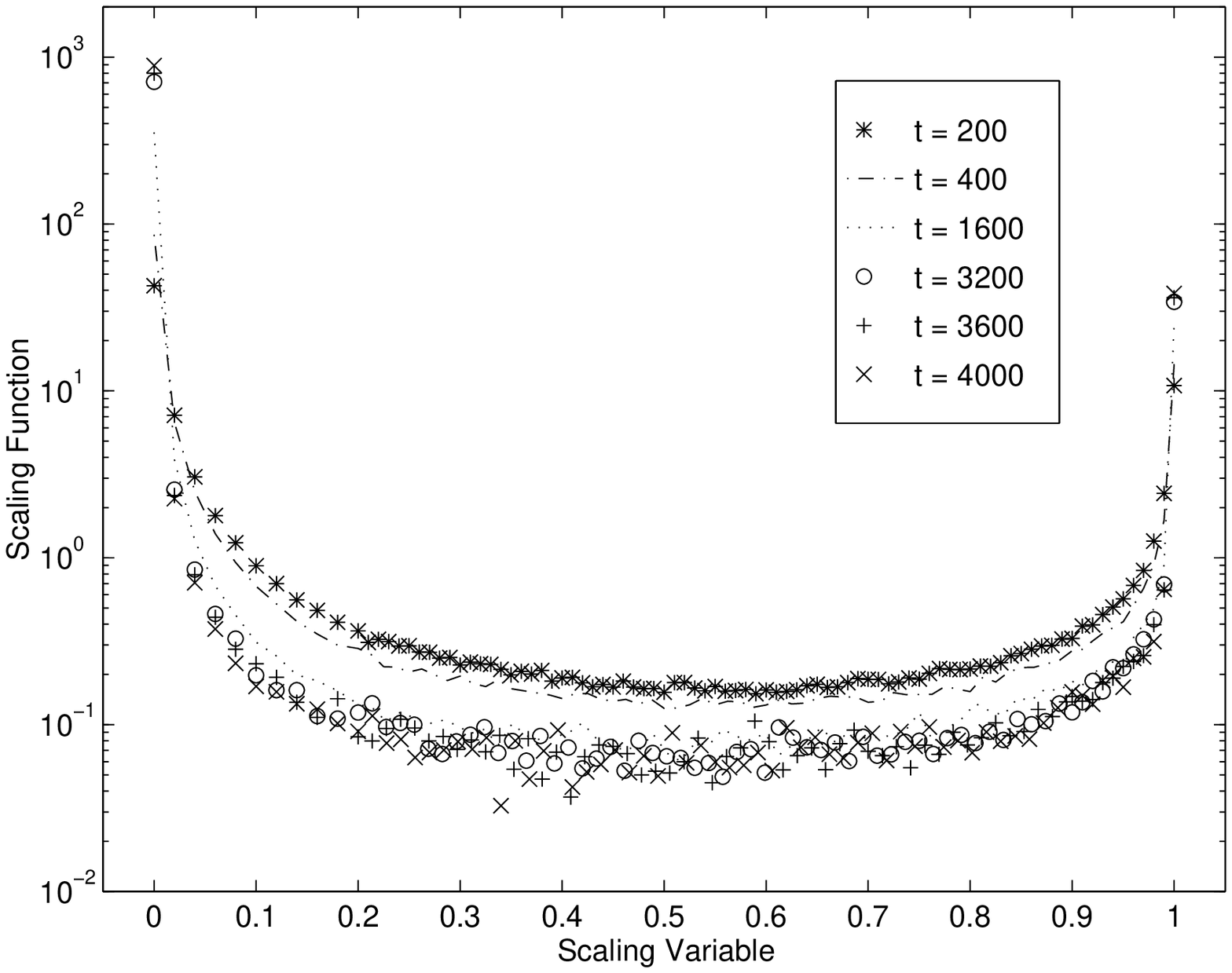}}
\caption{\protect The  life time distribution plotted against the scaling variable, for the
           zero temperature two dimensional Ising-Glauber coarsening
}
\end{figure}
\newpage
\begin{figure}
\centerline{
\psfig{figure=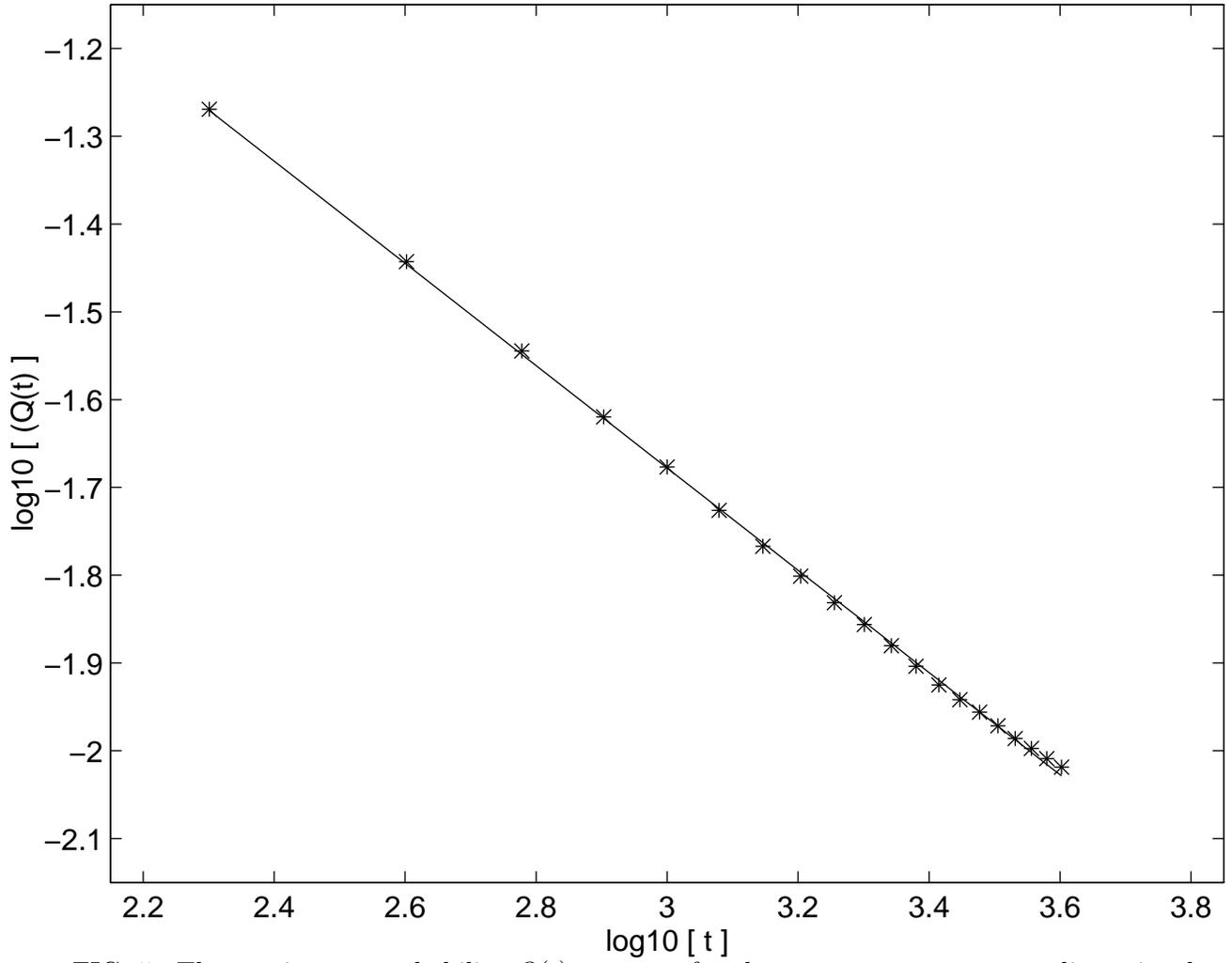}}
\caption{\protect The persistence probability $Q(t)$ {\it vs. } $t$ , for the
            zero temperature two dimensional Ising-Glauber coarsening.
            The solid line is the linear least square fit. The slope is
            $-0.5828$.} 

\end{figure}
\newpage
\begin{figure}
\centerline{
\psfig{figure=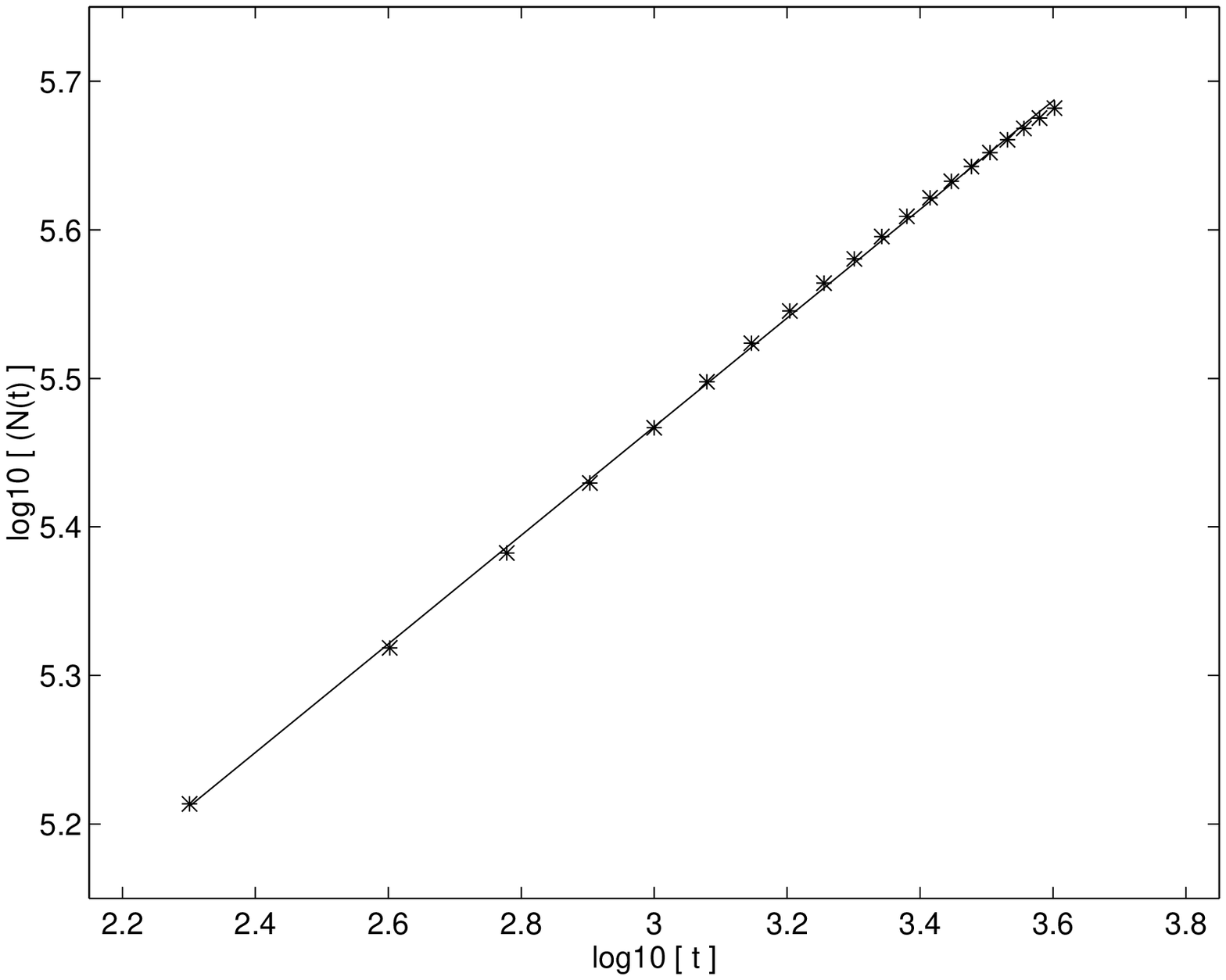}}
\caption{\protect The average number of lives per spin for the zero temperature
           two dimensional Ising-Glauber coarsening. The solid line
           is the linear least square fit. The slope is $0.3656$.
}
\end{figure}

\begin{thebibliography}{99}
\bibitem{ajb}
A. J. Bray, Adv. Phys., {\bf 43} 357 (1994).
\bibitem{Feller}
W. Feller, {\it Introduction to Probability theory abnd Its Application},
Vol. I (Chapter III), Third Edition, Wiley Eastern, New Delhi (1978).
\bibitem{Levy}
P. L\'evy, Composita Mathematica, {\bf 7}, 283 (1938), cited in
reference 2. 
\bibitem{snm}
S. N. Majumdar, Current Science (India), {\bf 77} 370 (1999).
\bibitem{bd}
B. Derrida, A. J. Bray and C. Godr\'eche, J. Phys. A{\bf 27}
L357 (1994);
B. Derrida, V. Hakim, and V. Pasquier, Phys. Rev. Lett., {\bf 75}
751 (1995);
B. Derrida, V. Hakim, and V. Pasquier, J. Stat. Phys., {\bf 85}, 763
(1996).
\bibitem{theory}
D. Stauffer, J. Phys. A{\bf 27} 5029 (1994);
P. L. Krapivsky, E. Ben-Naim and S. Redner, Phys. Rev. E{\bf 50}
2474 (1994);
A. J. Bray, B. Derrida and C. Godr\'eche, Europhys. Lett., {\bf 27}
175 (1994);
S. N. Majumdar and C. Sire, Phys. Rev. Lett., {\bf 77} 1420 (1996);
S. N. Majumdar, C. Sire, A. J. Bray and S. J. Cornell, Phys. Rev. Lett.,
{\bf 77} 2867 (1996);
B. Derrida, V. Hakim and R. Zeitak, Phys. Rev. Lett., {\bf 77} 2871 (1996);
A. Watson, Science {\bf 274} 919 (1996);
E. Ben-Naim, L. Frachebourg and P. L. Krapivsky, Phys. Rev. E {\bf 53} 
 3078 (1996);
L. Frachebourg, P. L. Krapivsky and S. Redner, Phys. Rev. E {\bf 55}
6684 (1997);
B. P. Lee and A. Rutinberg, Phys. Rev. Lett., {\bf 79} 4842 (1997);
S. N. Majumdar and A. J. Bray, Phys. Rev. Lett., {\bf 81} 2626 (1998).
\bibitem{experiment}
M. Marcos-Martin, D. Beysens, J-P Bouchaud, C. Godr\'eche and I. Yekutieli,
Physica D{\bf 214} 396 (1995);
B. Yurke, A. N. Pargellis, S. N. Majumdar and C. Sire, Phys. Rev. E {\bf 56}
 R40 (1997);
W. Y. Tam, R. Zeitak, K. Y. Szeto and J. Stavans, Phys. Rev. Lett.,
{\bf 78} 1588 (1997;
G. P. Wong, {\it Measurement of persistence in 1-
D diffusion}, physics/000248 (2000). 
\bibitem{extensions}
P. L. Krapivsky and E. Ben-Naim, Phys. Rev. E {\bf 56} 3788 (1998);
I. Doric and C. Godr\'eche, J. Phys. A {\bf 31} 5413 (1998);
D. S. Fisher, P. Le Doussal and C. Monthus, Phys. Rev. Lett., {\bf 80}
3539 (1998);
A. Baldassari, J. P. Bouchaud, I. Dornic, and C. Godr\"eche, Phys. Rev.
E{\bf 59} R20 (1999);
I. Dornic, A. Lemaitre, A. Baldassarri and H. Chate,
J. Phys. A: Math. Gen. {\bf 33} 7499 (2000).
\bibitem{smv}
V. Sridhar, M. C. Valsakumar and K. P. N. Murthy, 
{\it Modelling of coarsening dynamics employing 
distributions with power law tails and correlations}, under preparation (2000). 
\end{thebibliography}
\end{document}